\documentclass[pas,prb,twocolumn,showpacs,floatfix]{revtex4-1}

\usepackage{graphicx}
\usepackage{dcolumn}
\usepackage{bm}
\usepackage{amssymb}
\usepackage{hyperref}
\usepackage{amsmath}
\usepackage{mathrsfs}

\newcommand{\siesta}{\textsc{siesta~}}
\newcommand{\vasp}{\textsc{vasp~}}
\newcommand{\nemo}{\textsc{nemo~\oldstylenums{3}-d}}

\begin{document}

\title{Electronic properties of delta-doped phosphorus layers in silicon and germanium}

\author{J. S.~Smith}
\email{jackson.smith@rmit.edu.au}% http://www.researcherid.com/rid/I-1413-2013
\affiliation{Chemical and Quantum Physics, School of Applied Sciences, RMIT University, Melbourne VIC 3001, Australia}

\author{J. H.~Cole}
\email{jared.cole@rmit.edu.au}
\affiliation{Chemical and Quantum Physics, School of Applied Sciences, RMIT University, Melbourne VIC 3001, Australia}

\author{S. P.~Russo}
\email{salvy.russo@rmit.edu.au}
\affiliation{Chemical and Quantum Physics, School of Applied Sciences, RMIT University, Melbourne VIC 3001, Australia}

\date{\today}

\begin{abstract}
The Thomas-Fermi-Dirac (TFD) approximation and an $sp^{3}d^{5}s^{*}$ tight binding method were used to calculate the electronic properties of a $\delta$-doped phosphorus layer in silicon. This self-consistent model improves on the computational efficiency of ``more rigorous'' empirical tight binding and \textit{ab initio} density functional theory models without sacrificing the accuracy of these methods. The computational efficiency of the TFD model provides improved scalability for large multi-atom simulations, such as of nanoelectronic devices that have experimental interest. We also present the first theoretically calculated electronic properties of a $\delta$-doped phosphorus layer in germanium as an application of this TFD model.
\end{abstract}


\maketitle

\section{\label{sec:introduction}Introduction}

New \textit{in situ} phosphorus $\delta$-doping techniques in single crystal silicon and germanium are achieving unprecedented carrier concentrations inside highly confined layers ($\delta$-layers) of the host material.~\cite{Wilson2006a,Scappucci2012b} $\delta$-doped P layers in Si (Si:P) are ideally one monolayer (ML) thick and at very low temperatures exhibit electronic properties that are a combination of those of the dopant and the host material. Through advances in scanning tunnelling microscope lithography~\cite{Tucker1998a} and molecular beam epitaxy~\cite{OBrien2001a} a new generation of $\delta$-doped structures in Si is now being realised. These structures include: $\delta$-doped Si:P layers,~\cite{Shen2002a} quantum dots,~\cite{Fuechsle2010a,Fuhrer2009a} quantum dot arrays,\cite{Pok2007a} quantum wires~\cite{McKibbin2013a,Weber2012a} and a single electron transistor.~\cite{Fuechsle2012a}

Using \textit{in situ} doping techniques one in four atoms inside a Si(001) ML (1/4 ML) can be substituted with a P donor.~\cite{Wilson2006a} Highly doped planes of P atoms form layers, inside which the donor atoms are inherently disordered.~\cite{Wilson2006a} Because of the high doping densities inside these layers, we expect the donor electrons to behave similarly to an inhomogeneous electron gas. However, the majority of recent theoretical models describe the donor electrons using explicit donor atom configurations rather than an average donor electron density. Experimental verification of either type of theoretical model is currently unavailable as recent measurements of the Si:P $\delta$-layer band structure are inconclusive when compared to the theory.\cite{Miwa2013a} Presently, it is only through computational modelling that the electronic properties of Si:P $\delta$-layers can be quantitatively understood.

The most recent empirical and \textit{ab initio} models of Si:P $\delta$-layers are designed to maximise the completeness of the system's mathematical description rather than computational efficiency. Here, we show the efficacy of these models to be reproducible with a one-dimensional (1D) model of the donor electron potential using the Thomas-Fermi-Dirac (TFD) approximation~\cite{March1983,Parr1989} and a parametrisation of correlation effects.~\cite{Perdew1992a} We combine this description of the donor electrons with an empirical tight binding (TB) method,~\cite{Jancu1998a,Boykin2004b} which offers the scalability needed for large multi-atom simulations of nanoelectronic devices that have experimental interest.

One of the earliest models of a $\delta$-doped layer was the application of the Thomas-Fermi (TF) theory~\cite{Thomas1927a,Fermi1928a,March1983} to an n-type $\delta$-layer, in general.~\cite{Ioriatti1990a} Later, this same TF method was applied to p-type $\delta$-layers in GaAs and Si~\cite{Gaggero-Sager1998a} and, combined with TB in a study of n-type $\delta$-doped Si and B layers in GaAs.~\cite{Vlaev1998a} Transport properties in n-type $\delta$-layers in Si have also been studied using the TF method.~\cite{Rodriguez-Vargas2006a} Si:P $\delta$-layers were first modelled using a density functional theory (DFT) method with a planar Wannier orbital (PWO) basis,~\cite{Qian2005a} and later an antibonding orbital model~\cite{Chang1996a} with an $sp^{3}s^{*}$ TB method.~\cite{Cartoixa2005a} However, the $sp^{3}s^{*}$ basis set is too small to reproduce the experimentally predicted curvature of the $\vec{k}_{x}$ valley minima.~\cite{Jancu1998a} A self-consistent three-dimensional (3D) quantum well model~\cite{Rahman2007a} combined with an $sp^{3}d^{5}s^{*}$ TB method\cite{Jancu1998a,Boykin2004b} was therefore recently used in both studies of disorder~\cite{Lee2011a} and temperature dependence~\cite{Ryu2009a} in Si:P $\delta$-layers. These studies used the \nemo~package,~\cite{Klimeck2007a} which has also been applied to $\delta$-doped Si:P quantum wires.~\cite{Weber2012a,Ryu2013a} Complimentary DFT models of Si:P $\delta$-layers and quantum wires have been proposed using the \siesta and \vasp packages, with localised atomic orbital (LAO) bases~\cite{Carter2009a,Carter2011a,Drumm2013a,Budi2012a} and a planewave basis.~\cite{Drumm2013b} However, the applicability of these models to realistic device architectures is restricted by the $N^3$ scaling in calculation time associated with DFT. Historically, the effective mass theory (EMT) has also been used to study n-type $\delta$-doped layers in Si~\cite{Scolfaro1994a} and was recently put on a firm theoretical footing for Si:P $\delta$-layers.~\cite{Drumm2012a} Unfortunately, although computationally efficient, the EMT is only applicable to the lowest lying conduction valleys, which are assumed to be parabolic, and is therefore not suitable for modelling the higher conduction bands, which are important in the calculation of transport properties.

Here, we calculate the electronic properties of a Si:P $\delta$-layer using an $sp^{3}d^{5}s^{*}$ TB method and a self-consistent 1D (TB1D) model of the donor potential and, we present the first theoretically calculated electronic properties of a Ge:P $\delta$-layer. This also marks the first time that the TFD approximation has been combined with an $sp^{3}d^{5}s^{*}$ basis set. We report on the band structure, valley splitting, Fermi level, electronic density of states (eDOS) and the changes in these properties against ML doping density. The TB1D model is outlined in Section~\ref{sec:method}. The results of the model are discussed for a Si:P $\delta$-layer in Section~\ref{sec:benchmarking} and for a Ge:P $\delta$-layer in Section~\ref{sec:germanium}. We conclude in Section~\ref{sec:conclusion}.

\section{\label{sec:method}Method of calculation}
\subsection{\label{sec:level2}Hamiltonian}\label{sec:hamiltonian}

The electronic properties of bulk Si and Ge are described in the tight binding approximation. The Hamiltonian is defined using an $sp^{3}d^{5}s^{*}$ basis set~\cite{Jancu1998a} with 18 empirical parameters that correctly predict the bulk effective masses of Si and Ge.~\cite{Boykin2004b} The Hamiltonian is of the form,

\begin{equation}
     \hat{H}_{\rm{bulk}}=\sum\limits_{i,u}\varepsilon_{i}^{(u)}c_{i,u}^{\dagger}c_{i,u}^{\phantom{\dagger}}+\sum\limits_{i,j\ne i,u,v}t_{ij}^{(uv)}c_{i,u}^{\dagger}c_{j,v}^{\phantom{\dagger}}+\hat{H}_{\rm{b.c.}}\ ,
\end{equation}

where $c_{i,u}^{\dagger}$ and $c_{i,u}$ are creation and annihilation operators. The first term in $\hat{H}_{\rm{bulk}}$ describes the on-site energies \big($\varepsilon_{i}^{(u)}$\big) of an electron in orbital $u$ on atom $i$ and the second term the coupling \big($t_{ij}^{(uv)}$\big) between electron orbitals $u$ and $v$ on first nearest-neighbour atoms $i$ and $j$. The third term ($\hat{H}_{\rm{b.c.}}$) describes the boundary conditions which for periodic boundaries are of the form,

\begin{equation}
     \hat{H}_{\rm{b.c.}}=\sum\limits_{k,l\ne k,u,v}t_{kl}^{(uv)}c_{k,u}^{\dagger}c_{l,v}^{\phantom{\dagger}}\ ,
\end{equation}

where $k$ and $l$ are first nearest-neighbour atoms via the periodic boundaries of the supercell. The Hamiltonian terms $\varepsilon_{i}^{(u)}$ and $t_{ij}^{(uv)}$ are expressed through TB parameters.~\cite{Slater1954a} These parameters are found by an optimisation procedure which fits to experimentally measured electronic properties: band energies and effective masses at high symmetry points in the Brillouin zone (BZ).~\cite{Klimeck2000a} We do not carry out such a procedure here, and instead use previously published TB parameters for Si and Ge.~\cite{Boykin2004b}

The $sp^{3}d^{5}s^{*}$ basis set comprises 20 hydrogenic orbitals for the description of the bulk Si and Ge band structure. In the $\delta$-layer system, the P donor electrons occupy conduction states that are non-degenerate~\cite{Miwa2013a} and we are therefore able to ignore electron spin and halve the size of the TB basis set from 20 to 10 orbitals. This results in a Hamiltonian matrix of order $4800\times 4800$. To calculate the electronic properties of the structure the Hamiltonian matrix is diagonalised for all $\vec{k}$-points in the first irreducible Brillouin zone (IBZ).~\cite{Ramiarez1986a} The speed of this diagonalisation is directly proportional to the order of the Hamiltonian matrix and the number of $\vec{k}$-points in the first two-dimensional (2D) BZ or $\vec{k}$-point grid. We use an effective $\vec{k}$-point grid of $120{\times}120{\times}1$ to converge the Fermi level to within 1~meV.

To simulate the $\delta$-layer we use a primitive tetragonal (tP) supercell which has dimensions $(a\times a\times 60a)$ nm, where $a$ is the lattice constant of the host material. This is equivalent to 120 MLs of bulk ``cladding'' in the positive and negative $z$-directions perpendicular to the $\delta$-layer. Periodic boundary conditions are applied in the $x$ and $y$ directions, and also in the $z$ direction after the bulk cladding. Using 120 MLs of bulk cladding the energies of the conduction states of interest are converged to within 1~meV. We do not geometry optimise the bulk Si or Ge as previous DFT investigations have shown that the effect of the P atoms on the positions of the bulk Si atoms is negligible.~\cite{Drumm2013b,Carter2009a} However, we expect the incorporation of P atoms into bulk Ge to have a larger effect on the positions of the Ge atoms because the Ge nucleus is much larger than the Si and P nuclei. Although we do not geometry optimise the bulk Ge here, we note that this approximation merits further investigation.

\begingroup
\begin{table}[t!]
\begin{ruledtabular}
     \caption{Parameters used in the calculations and references in square brackets. $m_{0}$ is the free electron mass.}
     \begin{tabular*}{\columnwidth}{@{\extracolsep{\fill} } l c c c c }
           & $\epsilon_{r}$ & $m_{l}/m_{0}$ & $m_{t}/m_{0}$ & $a$~(\AA) \\
          \hline
          Si & 11.4~[\onlinecite{Faulkner1969a}] & 0.9163~[\onlinecite{Hensel1965a}] & 0.1905~[\onlinecite{Hensel1965a}] & 5.430~[\onlinecite{Jancu1998a}] \\
          Ge & 15.36~[\onlinecite{Faulkner1969a}] & 1.588~[\onlinecite{Levinger1961a}] & 0.08152~[\onlinecite{Levinger1961a}] & 5.6563~[\onlinecite{Jancu1998a}]
     \label{parameters}
     \end{tabular*}
\end{ruledtabular}
\end{table}
\endgroup

\subsection{\label{sec:potential}Donor electron potential}

The donor electrons are described by a 1D potential that is added to the diagonal terms of the Hamiltonian matrix.~\cite{Graf1995a,Vlaev1998a} The TFD approximation and an appropriate parametrisation of correlation effects are used to calculate the donor potential. In this calculation the doping plane is approximated by a sheet of vanishing thickness with a constant areal doping density, and the validity of the local density approximation is assumed throughout. The TF and Dirac exchange formalism are exact in the high density limit. The TF approximation for $\delta$-doped semiconductor structures has been shown to produce meaningful results in the $10^{12}$--$10^{13}~\rm{cm}^{-2}$ electron density range~\cite{Ioriatti1990a} (one to two orders of magnitude lower than the nominal 1.7$\times$10$^{14}~\rm{cm}^{-2}$ considered herein). A full derivation of the TF theory for $\delta$-doped semiconductor structures is presented in the Appendix. For a 2D $\delta$-doped semiconductor structure in the high density limit, the electrostatic potential in the TF approximation is written as
\begin{equation}
     V_{\rm{TF}}(z)=-\frac{\alpha^{2}}{(\alpha|z|+z_{0})^{4}}\ ,
\end{equation}
with
\begin{equation}
     \alpha=\frac{(2\bar{m})^{3/2}e^{2}\nu}{60\pi^{2}\epsilon_{r}\epsilon_{0}\hbar^{3}}
\end{equation}
and
\begin{equation}
     z_{0}=\left(\frac{8\alpha^{3}\epsilon_{r}\epsilon_{0}}{e^{2}n_{D}}\right)^{1/5}\ ,
\end{equation}
where $z$ is distance perpendicular to the $\delta$-layer, $\nu$ is the number of equivalent conduction valleys, $\epsilon_{r}$ is the static dielectric constant, $\bar{m}$ is the geometric average of the longitudinal ($m_{l}$) and transverse ($m_{t}$) effective masses~\cite{Drumm2012a,Rodriguez-Vargas2006a,Scolfaro1994a,Brinkman1973a} ($\bar{m}=m_{t}^{2/3}m_{l}^{1/3}$) and $n_{D}$ is the areal donor density. The parameters in Table~\ref{parameters} are used throughout.

To the electrostatic potential is added an exchange potential ($V_{\rm{X}}$) derived from the Dirac exchange energy functional~\cite{Dirac1930a} and a correlation potential ($V_{\rm{C}}$). The exchange potential is written as
\begin{equation}
     V_{\rm{X}}(z)=-\frac{\bar{m}e^{4}}{(4\pi\epsilon_{r}\epsilon_{0}\hbar)^{2}}\left(\frac{9}{4\pi^{2}}\right)^{1/3}\frac{1}{r_{s}(z)}\ ,
\end{equation}
with
\begin{equation}
     r_{s}(z)=\left(\frac{4\pi\bar{a_{0}}^{3}n(z)}{3}\right)^{-1/3}
\end{equation}
and
\begin{equation}
     \bar{a_{0}}=\frac{4\pi\epsilon_{r}\epsilon_{0}\hbar^{2}}{\bar{m}e^{2}}
     \label{bohr_radius}
\end{equation}
where $r_{s}$ is the Wigner-Seitz radius, $\bar{a_{0}}$ is the effective Bohr radius ($\bar{a_{0}}=1.96~\rm{nm}$ for Si, and $\bar{a_{0}}=3.80~\rm{nm}$ for Ge) and $n(z)$ is the electron density (see Appendix). The correlation energy is approximated using the correlation functional parametrised by Perdew~\&~Wang~\cite{Perdew1992a} for systems having zero spin polarisation. In the high density limit, this correlation potential is written as
\begin{align}
     \nonumber V_{\rm{C}}(z)=&-2A\left\{\ln\left[1+\frac{1}{2Af}\right]\left(1+\frac{2\alpha_{1}r_{s}(z)}{3}\right)\right.\\
     &\left.- \ \frac{f'}{f\left(1+2Af\right)}\left(\frac{r_{s}(z)(1+\alpha_{1}r_{s}(z))}{3}\right)\right\}\ ,
\end{align}
with
\begin{equation}
     f=\beta_{1}r_{s}^{1/2}+\beta_{2}r_{s}+\beta_{3}r_{s}^{3/2}+\beta_{4}r_{s}^{2}
\end{equation}
and
\begin{equation}
     f'=\frac{\beta_{1}}{2}r_{s}^{-1/2}+\beta_{2}+\frac{3\beta_{3}}{2}r_{s}^{1/2}+2\beta_{4}r_{s}\ ,
\end{equation}
where $A=0.031091$, $\alpha_{1}=0.21370$, $\beta_{1}=7.5957$, $\beta_{2}=3.5876$, $\beta_{3}=1.6382$, and $\beta_{4}=0.49294$. Following the methods of Refs.~\onlinecite{Drumm2012a,Rodriguez-Vargas2006a,Scolfaro1994a} we do not consider the anisotropy of the effective mass term in the calculation of the exchange and correlation (XC) potentials.

Finally, the donor potential is written as
\begin{equation}
     V(z)=V_{\rm{TF}}(z)+V_{\rm{X}}(z)+V_{\rm{C}}(z)\ ,
\end{equation}
which in matrix form is
\begin{equation}
     \hat{V}=\sum\limits_{i}V_{i}^{\phantom{\dagger}}c_{i}^{\dagger}c_{i}^{\phantom{\dagger}}\ ,
\end{equation}
where $V_{i}$ is the discrete form of $V(z)$ for the $z$ coordinate. The Hamiltonian describing the $\delta$-layer system is therefore,
\begin{equation}
     \hat{H}=\hat{H}_{\rm{bulk}}+\hat{V}\ .
\end{equation}
Here, the donor potential is treated as an external potential and added to only the diagonal or on-site energy terms of the Hamiltonian matrix:~\cite{Graf1995a,Vlaev1998a} an approximation we believe is validated by the results of the TB1D model as demonstrated in Section~\ref{sec:benchmarking}.

\subsection{\label{sec:fermi}Fermi level pinning}

Recent measurements of the band structure of a Si:P $\delta$-layer show at least one occupied state in the conduction band (CB).~\cite{Miwa2013a} Because the $\delta$-layer exhibits metallic characteristics at low temperatures, the Fermi level must be solved for iteratively. In this work, the number of occupied conduction states is increased from zero until the charge neutrality condition is satisfied, \textit{i.e.} inside a finite area, the number of electrons is equal to the number of donor atoms (each donor atom contributing one donor electron). The energy at which charge neutrality is achieved is defined as the energy of the Fermi level. The total number of electrons at energy $E$ can be written as
\begin{equation}
     N(E)=s\ f(E)\ Z(E)\ ,
\end{equation}
where $s$ is the spin degeneracy, $Z(E)$ is the eDOS evaluated over the first 2D BZ, and $f(E)$ is the Fermi-Dirac distribution function. For our calculations $s$ is equal to one, therefore,
\begin{equation}
     N(E)=Z(E)\left(1+\exp{\bigg[\frac{E-E_{\rm{F}}}{k_{\rm{B}}T}\bigg]}\right)^{-1}\ ,
     \label{eq:fermi_level}
\end{equation}
where $E_{\rm{F}}$ is the Fermi level, $k_{\rm{B}}$ is Boltzmann's constant, and $T$ is temperature. The area of the first 2D BZ is converted to a real space area and the number of donors is calculated from the areal doping density. $E_{\rm{F}}$ is then solved for iteratively by evaluating Eq.~\ref{eq:fermi_level} at energies greater than the bulk valence band minimum until charge neutrality is achieved.

\section{\label{sec:benchmarking}Analysis of the model for a Si:P $\delta$-layer}

\begin{figure}[b!]
    \centering
    \includegraphics{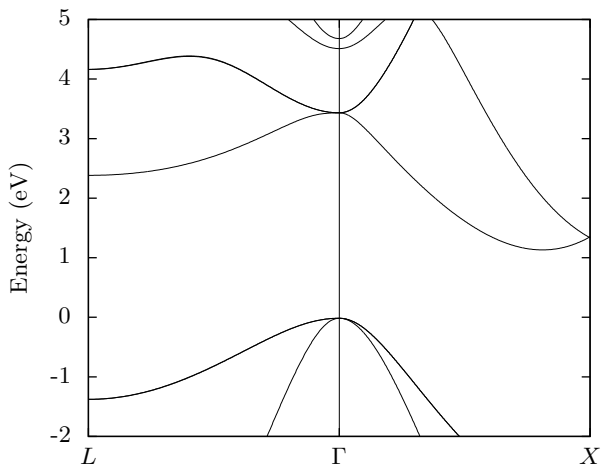}
    \caption{Band structure of bulk Si calculated using an $sp^{3}d^{5}s^{*}$ TB method.~\cite{Jancu1998a,Boykin2004b} The bands are plotted in the FCC BZ.}
    \label{bulk_si_bands}
\end{figure}

\begin{figure}[t!]
    \centering
    \includegraphics{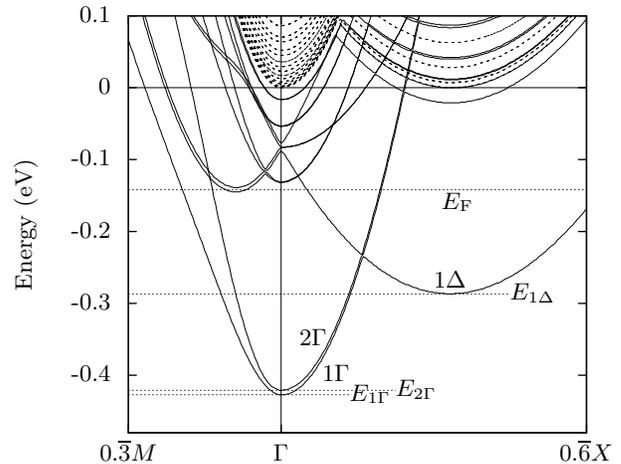}
    \caption{Band structure (12 lowest conduction bands only) of a Si:P $\delta$-layer (1/4~ML) [solid lines], the energy minima of the lowest three conduction bands [dotted lines], and the band structure (CB only) of bulk Si [dashed lines].~\cite{Jancu1998a,Boykin2004b} The bands are plotted in the tP BZ and the energy offset has been chosen to set the bulk Si CB minimum to zero.}
    \label{si_bands}
\end{figure}

Si is an indirect band gap semiconductor and has a six-fold degenerate CB minimum at $\pm0.81A$ in Figure~\ref{bulk_si_bands}, where $A=X,Y,Z$ are points of high symmetry in the face-centred cubic (FCC) BZ.~\cite{Bradley1972a} Figure~\ref{bulk_si_bands} shows the band structure of bulk Si plotted in the FCC BZ on the path of high symmetry: $L$ to $\Gamma$ to $X$. Figure~\ref{si_bands} shows the band structure (CB only) of a Si:P $\delta$-layer plotted in the tP BZ on the path of high symmetry: $0.\overline{3}M$ to $\Gamma$ to $0.\overline{6}X$. The path $\Gamma$ to $Y$ is equivalent to the path $\Gamma$ to $X$ by the four-fold symmetry of the Si(001) doping plane and therefore not shown. The path $\Gamma$ to $Z$ is affected by zone-folding~\cite{Boykin2005a,Drumm2013b} because of the height of the tP supercell in the $z$ direction and therefore also not shown. The change from an FCC unit cell to a tP supercell projects the  conduction valleys in the $\pm z$ directions to $\Gamma$. These conduction bands are labelled $1\Gamma$ and $2\Gamma$ in Figure~\ref{si_bands}. The vertical confinement of electrons by the donor potential shifts these conduction valleys into the bulk band gap region and splits their energy minima by the $1\Gamma/2\Gamma$ valley splitting.~\cite{Carter2009a} The change to a tP supercell also folds the bands in the $\pm x$ and $\pm y$ directions. The $\pm0.81X$ and $\pm0.81Y$ valley minima of bulk Si are folded to $\pm0.37X$ and $\pm0.37Y$ respectively. The $+X$ valley is labelled as $1\Delta$ in Figure~\ref{si_bands}. The projection of the Si conduction valleys is illustrated in Ref.~\onlinecite{Qian2005a}~[Fig.~1].

The 1D confinement of the donor electrons in the $z$ direction shifts the lowest conduction states of bulk Si into the bulk band gap region.  Figure~\ref{si_bands} shows three conduction valleys to be partially occupied, or partly below the Fermi level ($E_{\rm{F}}$): the $1\Gamma$, $2\Gamma$ and $1\Delta$ bands. The Si:P $\delta$-layer is metallic; the CB minimum ($E_{1\Gamma}$) is 285~meV below $E_{\rm{F}}$. $E_{1\Gamma}$ is $427$~meV below the CB minimum of bulk Si and, in Table \ref{si_bands_minima}, agrees with the results of all models (excluding DFT1D (LAO)~\cite{Carter2011a} and DFT3D~\cite{Drumm2013a}) to within 28~meV. The minimum of the $2\Gamma$ band ($E_{2\Gamma}$) compares equally well to DFT1D (PWO)~\cite{Qian2005a} and is within 5~meV of EMT1D.~\cite{Drumm2012a} Interestingly, the $1\Delta$ valley minimum ($E_{1\Delta}$) of TB1D is 30--38~meV lower than that predicted by the three closest models in Table~\ref{si_bands_minima}. In general, there is good agreement between the literature and TB1D. Contrastingly, values for $E_{\rm{F}}$ differ significantly between the models. This disparity can be explained by the variety in the exact method and $\vec{k}$-point grid used to solve for the Fermi level between each of the models. To calculate $E_{\rm{F}}$, we use a similar procedure to TB3D~\cite{Ryu2013a} and although TB1D predicts $E_{\rm{F}}$ to be 27~meV lower than TB3D,~\cite{Ryu2013a} this discrepancy is explained by the 26~meV difference between the values for $E_{1\Gamma}$. The calculated binding energies ($E_{\rm{F}}-E_{1\Gamma}$) of TB3D~\cite{Ryu2013a} and TB1D agree within 1~meV (see Table~\ref{valley_splitting}). 

\begingroup
\begin{table}[t!]
\begin{ruledtabular}
     \caption{Energies of conduction valley minima and Fermi levels (meV) at low temperature for a variety of Si:P $\delta$-layer models (1/4~ML). The models are arranged in order of increasing magnitude of $E_{1\Gamma}$. This work is highlighted in bold. Uncertainty of $\pm$3~meV is labelled by $\dagger$.}
     \begin{tabular*}{\columnwidth}{@{\extracolsep{\fill} } l c c c c }
          Model (basis set) & $E_{1\Gamma}$ & $E_{2\Gamma}$ & $E_{1\Delta}$ & $E_{\rm{F}}$ \\
          \hline
          DFT1D (LAO)~\cite{Carter2011a} $\dagger$ &  -296 & -288 & -165 & -72 \\
          DFT3D (LAO)~\cite{Drumm2013b} & -369 & -269 & -68 & -23 \\ 
          TB3D ($sp^{3}d^{5}s^{*}$)~\cite{Ryu2013a} & -401 & -375 & -249 & -115 \\
          DFT1D (PWO)~\cite{Qian2005a} $\dagger$ & -419 & -394 & -250 & -99 \\
          \textbf{TB1D ($sp^{3}d^{5}s^{*}$)} & \textbf{-427} & \textbf{-421} & \textbf{-287} & \textbf{-142}  \\
          EMT1D~\cite{Drumm2012a} $\dagger$ & -445 & -426 & -257 & N/A
     \label{si_bands_minima}
     \end{tabular*}
\end{ruledtabular}
\end{table}
\endgroup

\begin{figure}[t!]
    \centering
    \includegraphics{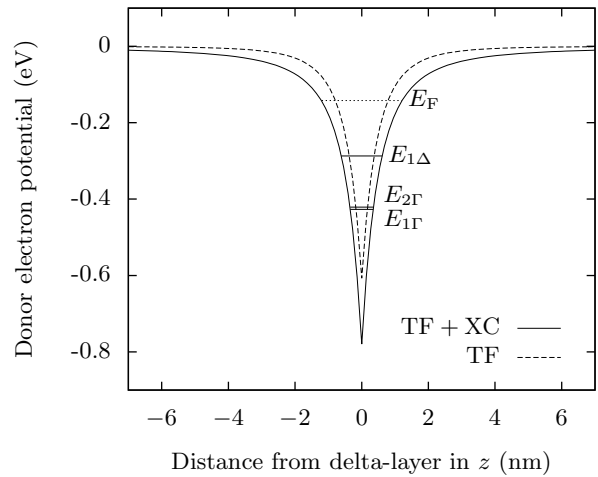}
    \caption{Donor potential for a Si:P $\delta$-layer (1/4~ML) with [solid line] and without [dotted line] XC effects. The Fermi level and energies of the conduction valley minima are shown as energy levels inside the potential well.}
    \label{fig:donor_potential}
\end{figure}

Similarly, the greatest contributing factor to the discrepancies in the values of the conduction valley minima between the models in Table~\ref{si_bands_minima} can be explained by the exact methods used to estimate the XC energy. An examination of the conduction valley energy minima of DFT1D (PWO)~\cite{Qian2005a} shows a 10$\pm$3~meV shift in $E_{1\Gamma}$ when XC and short-range effects are added to the model, compared to a -34~meV shift in the $E_{1\Gamma}$ for DFT3D~\cite{Ryu2013a} when XC effects are included (see Ref.~\onlinecite{Ryu2009a}). These XC energy corrections are relatively small compared to the -130~meV shift in $E_{1\Gamma}$ that corresponds to including XC effects in TB1D (see Figure~\ref{fig:donor_potential} for a comparison of the donor potentials with and without XC). To approximate the XC energy correction to the semiclassical component of the donor potential, we use the same procedure as EMT1D. In Table~\ref{si_bands_minima}, the values of the conduction valley minima for TB1D agree most strongly with EMT1D. The method used by DFT1D (PWO)~\cite{Qian2005a} to estimate the XC energy correction is available and in future could be used to test the validity of the two different approximations to XC. Unfortunately, the details of the exact implementation of XC in TB3D~\cite{Ryu2013a} are not available. However, the lack of quantitative experimental results for the Si:P $\delta$-layer system means verification of any of the approximations to XC effects is currently impractical.

Figure~\ref{fig:donor_potential} shows the TF donor potential with and without XC. The energies of the conduction valley minima are shown as energy levels inside the potential well. XC effects increase the magnitude of the TF potential. We find exchange effects dominate over correlation effects at the nominal 1/4~ML doping density. The exchange potential shifts the energies of the CB minimum by -125~meV, where as the correlation potential shifts the energies by only -5~meV. We note that the system is therefore adequately described by the TFD donor potential and that the correlation potential could be ignored for simplicity.

The degeneracy of the Si CB minimum results in a valley splitting between the $1\Gamma$ and $2\Gamma$ bands, named the $1\Gamma/2\Gamma$ valley splitting~\cite{Carter2011a} (equal to $E_{2}-E_{1}$). Understanding this energy splitting is important in the design of ``few-electron'' quantum electronics in silicon~\cite{Rahman2011a} and modelling transport properties in these systems at low voltage biases.~\cite{Fuechsle2010a} Table~\ref{valley_splitting} shows the values of $E_{2}-E_{1}$ calculated by a variety of Si:P $\delta$-layer models. The $1\Gamma/2\Gamma$ valley splitting calculated by TB1D is 6~meV and agrees with that of DFT1D (LAO)~\cite{Carter2011a} to within 2$\pm$6~meV. There is a difference of 20~meV between the valley splitting predicted by TB3D~\cite{Ryu2013a} and TB1D. There are also differences of 14$\pm$6 and 21$\pm$6~meV between the valley splitting predicted by EMT1D~\cite{Drumm2012a} and DFT1D (PWO)~\cite{Qian2005a} respectively and TB1D. These discrepancies can be explained by the confinement of electrons in the $z$ direction.~\cite{Carter2011a,Drumm2012a} The valley splitting is inversely proportional to the spatial extent of the donors perpendicular to the doping plane.~\cite{Budi2012a} There are two characterisitcs of the donor potential that affect the confinement of the donor electrons. These are the rate of decay of the gradient of the potential in the $z$ direction and the depth of the potential well. A comparison of the donor potentials and valley splittings between the models is consistent with this statement. The donor potential of DFT1D (PWO)~\cite{Qian2005a} (see Ref.~\onlinecite{Qian2005a} [Fig. 3]) compares well to Figure~\ref{fig:donor_potential}. The DFT1D (PWO)~\cite{Qian2005a} donor potential has a similar depth ($\sim$0.75~eV) to the TB1D potential (0.78 eV) however the rate of decay of the gradient of the potential in the $z$ direction is larger for DFT1D (PWO)~\cite{Qian2005a} (decaying to zero by $\sim$4~nm). The larger rate of decay results in stronger confinement of the electrons in the $z$ direction and a larger valley splitting (26$\pm$6~meV). The rate of decay of the gradient of the potential in the $z$ direction and the depth of the potential are sensitive to ML doping density (see Table~\ref{density_decay}), donor atom configuration~\cite{Carter2011a,Lee2011a} and basis set size.~\cite{Drumm2013b} We note that including XC effects in TB1D and DFT1D (PWO)~\cite{Qian2005a} and short-range effects in DFT1D (PWO)~\cite{Qian2005a} does not affect the magnitude of the valley splitting. And, that when a large basis set is used effects of basis set size become relatively minor.~\cite{Drumm2012a} Therefore, it is the ML doping density which controls the size of the $1\Gamma/2\Gamma$ valley splitting and it is understanding the proportionality between these properties which is important for engineering valley splitting in Si:P $\delta$-layer systems.

\begingroup
\begin{table}[t!]
\begin{ruledtabular}
     \caption{Values of valley splitting and binding energy (meV) at low temperature for for a variety of Si:P $\delta$-layer (1/4~ML) models and experiment. The models are arranged in order of increasing $E_{\rm{F}}-E_{1\Gamma}$. This work is highlighted in bold. Uncertainty of $\pm$3~meV is labelled by $\dagger$.}
     \begin{tabular*}{\columnwidth}{@{\extracolsep{\fill} } l c c}                        
          Model (basis set) & $E_{2\Gamma}-E_{1\Gamma}$ & $E_{\rm{F}}-E_{1\Gamma}$ \\
          \hline
          DFT1D (LAO)~\cite{Carter2011a} $\dagger$ & 8 & 224 \\
          Experiment~\cite{Miwa2013a} ($T=100K$) & N/A & $\sim$270~~ \\ 
          \textbf{TB1D ($sp^{3}d^{5}s^{*}$)} & \textbf{6} & \textbf{285}  \\
          TB3D ($sp^{3}d^{5}s^{*}$)~\cite{Ryu2013a} & 26 & 286  \\
          DFT1D (PWO)~\cite{Qian2005a} $\dagger$ & 25 & 320 \\
          DFT3D (LAO)~\cite{Drumm2013b} &  100 & 346 \\
          EMT1D~\cite{Drumm2012a} & 19 & 470$^\dagger$
     \label{valley_splitting}
     \end{tabular*}
\end{ruledtabular}
\end{table}
\endgroup

Recently published experimental measurements of the Si:P $\delta$-layer band structure performed at $T=100~K$ confirm the existence of at least one occupied state at $\Gamma$.~\cite{Miwa2013a} This state, named the $\delta$-state, has a binding energy ($E_{\rm{F}}-E_{1}$) of $\sim$190~meV at $T=300~K$ and $\sim$270~meV at $T=100~K$. A comparison to the theory is difficult as all Si:P $\delta$-layer models explicitly assume lower temperatures, in the neighbourhood of $T=4~K$. However, it is obvious from Table~\ref{valley_splitting} that the experimental binding energy of $\sim$270~meV at $T=100~K$ compares well to the value of 285~meV calculated by TB1D. At lower experimental temperatures we expect the binding energy of the $\delta$-state to continue to increase as thermal excitations are reduced, increasing the confinement of the donor electrons and decreasing the electron screening of the $\delta$-state by intrinsic charge carriers. It is not obvious from the experimental measurements whether there is a valley splitting at $\Gamma$. However, the energy resolution of the experimental band structure measurement is low and energy splittings of less than $\sim$70~meV are not resolvable. Ref.~\onlinecite{Miwa2013a} suggests that the $1\Gamma/2\Gamma$ valley splitting is either small and not able to be resolved or large (\textgreater150~meV) and significantly underestimated by the theoretical models available in the literature. In Figure~2(b) of Ref.~\onlinecite{Miwa2013a} the $1\Gamma$, $2\Gamma$ and $1\Delta$ conduction bands calculated in Ref.~\onlinecite{Carter2011a} [Figure~10(d)] are overlaid to show the disparity between theory and experiment. However, Ref.~\onlinecite{Carter2011a} has been shown to overestimate the $1\Gamma/2\Gamma$ valley splitting by $\sim$50\%.~\cite{Drumm2013b} We expect the experimental valley splitting to increase with stronger confinement at lower temperatures and therefore suggest that a small valley splitting (\textless70~meV) is more likely. This is consistent with the Si:P $\delta$-layer models in the Table~\ref{valley_splitting} (excluding DFT3D~\cite{Drumm2012a}).

\begingroup
\begin{table}[b!]
\begin{ruledtabular}
     \caption{Energies of the CB minimum, valley splitting, Fermi level and binding energy (meV) of a Si:P $\delta$-layer at low temperature for a range of ML doping densities calculated by TB1D and using DFT1D (PWO)~\cite{Qian2005a} for comparison.}
     \begin{tabular*}{\columnwidth}{@{\extracolsep{\fill} } l c c c c c c }
          ML doping density & 1/4 & 1/8 & 1/16 & 1/32 & 1/64 & 1/128 \\                       
          $n_{D}~(10^{14}~\rm{cm}^{-2})$ & 1.696 & 0.847 & 0.424 & 0.212 & 0.106 & 0.053 \\
          \hline
          $E_{1\Gamma}$ & -427 & -252 & -152 & -95 & -59 & -38 \\
          $E_{2\Gamma}-E_{1\Gamma}$ & 6 & 2 & 1 & 1 & 0 & 0 \\
          $E_{\rm{F}}$ & -142 & -90 & -60 & -43 & -29 & -20 \\
          $E_{\rm{F}}$ (DFT1D~\cite{Qian2005a}) & -111 & N/A & -62 & N/A & -36 & N/A \\
          $E_{F}-E_{1\Gamma}$ & 285 & 162 & 92 & 52 & 30 & 18
     \label{density_decay}
     \end{tabular*}
\end{ruledtabular}
\end{table}
\endgroup

Table~\ref{density_decay} shows the sensitivity of CB minimum, valley splitting, Fermi level and binding energy to changes in ML doping density. $E_{1\Gamma}$ increases from -427~meV at the nominal doping density of 1/4~ML ($1.696\times10^{14}~\rm{cm}^{-2}$) to -38~meV at a density of 1/128 ML ($5.3\times10^{12}~\rm{cm}^{-2}$). This shows that at a doping density as low as 1/128~ML, $E_{1\Gamma}$ can still be resolved from the CB minimum of bulk Si. However, we note that the TF donor potential is most accurate at high doping densities~\cite{Ioriatti1990a} (\textgreater$10^{13}~\rm{cm}^{-2}$) or at ML doping densities greater than 1/64~ML. In the range of doping densities greater than 1/64~ML, $E_{1\Gamma}$ and $E_{\rm{F}}$ show a linear response to changes in ML doping density. This trend agrees with DFT1D (PWO),~\cite{Qian2005a} as shown in Table~\ref{density_decay}. The strongest agreement between the Fermi levels of the two models is at a doping density of 1/16~ML at which they agree to within 1~meV. A consequence of this behaviour is that the binding energy ($E_{F}-E_{1\Gamma}$) also shows a linear response to changes in areal doping density. The TB1D model calculates a binding energy ($E_{\rm{F}}-E_{1\Gamma}$) of 52~meV at a doping density of 1/32~ML which is close to the experimental Si:P single donor level of 45.6~meV.~\cite{Ramdas1981a} This suggests that at a doping density of $\sim$1/32~ML, the separation of the P atoms is sufficient to recover the physics of a single P donor in Si. An areal doping density of 1/32~ML corresponds to an average donor separation of 2.17~nm. This agrees with the effective Si:P Bohr radius in the effective mass approximation (1.96~nm) calculated using Eq.~\ref{bohr_radius} and the parameters in Table~\ref{parameters}. It also agrees with Ref.~\onlinecite{Wellard2005a} [Figure~1, $\eta=5.8$] which shows the probability density of a single P donor in Si to have decayed significantly by $\sim$2~nm of bulk Si cladding. This suggests the TB1D model is applicable at ML doping densities as low as $\sim$$2$$\times$$10^{13}~\rm{cm}^{-2}$.
 
\section{\label{sec:germanium}Application of the model to a Ge:P $\delta$-layer}

\begin{figure}[b!]
    \centering
    \includegraphics{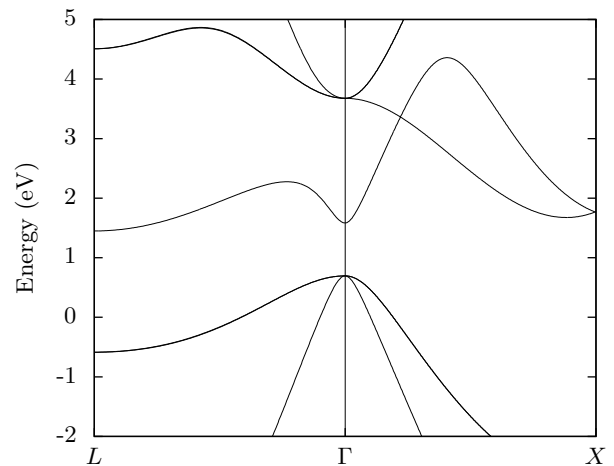}
    \caption{Band structure of bulk Ge calculated using an $sp^{3}d^{5}s^{*}$ TB method.~\cite{Jancu1998a,Boykin2004b} The bands are plotted in the FCC BZ.}
    \label{ge_bulk}
\end{figure}

\begin{figure*}[th!]
    \centering
    \includegraphics{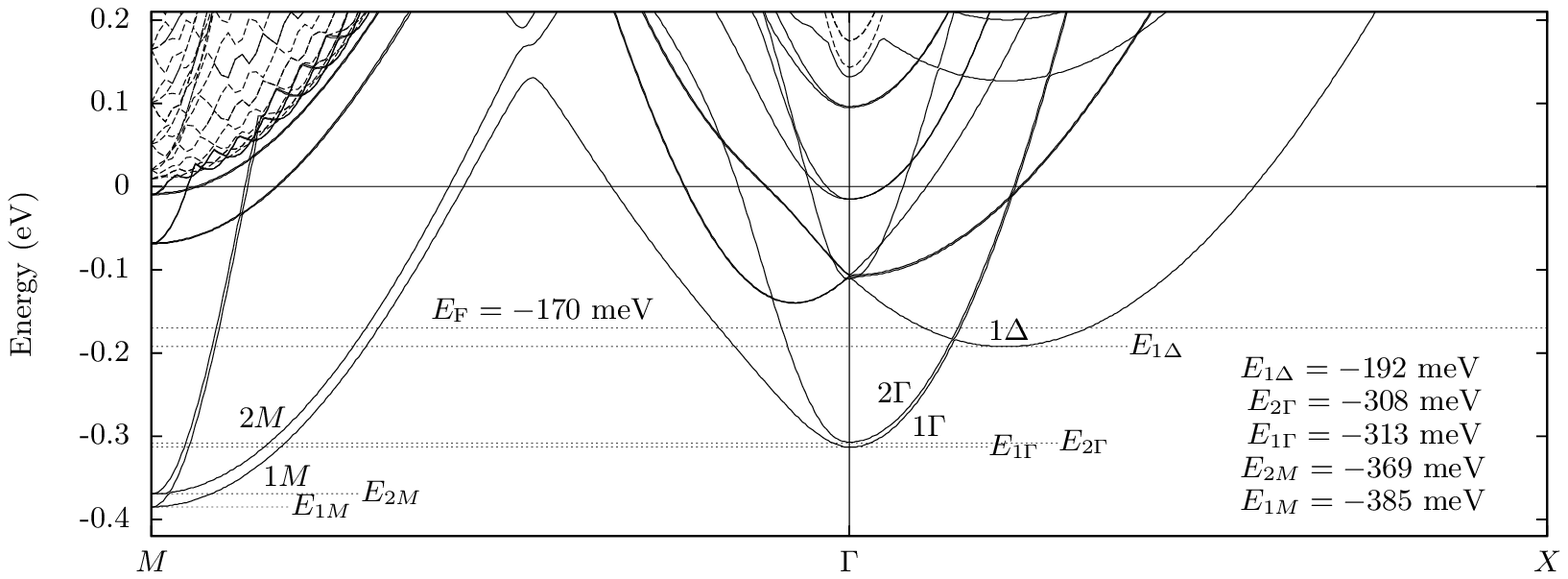}
    \caption{Band structure (12 lowest conduction bands only) of a Ge:P $\delta$-layer (1/4~ML) [solid lines], the energy minima of the lowest five conduction bands [dotted lines], and the band structure (CB only) of bulk Ge [dashed lines].~\cite{Jancu1998a,Boykin2004b} The bands are plotted in the tP BZ and the energy offset has been chosen to set the bulk Ge CB minimum to zero.}
    \label{ge_bands}
\end{figure*}

\begin{figure}[th!]
    \centering
    \includegraphics{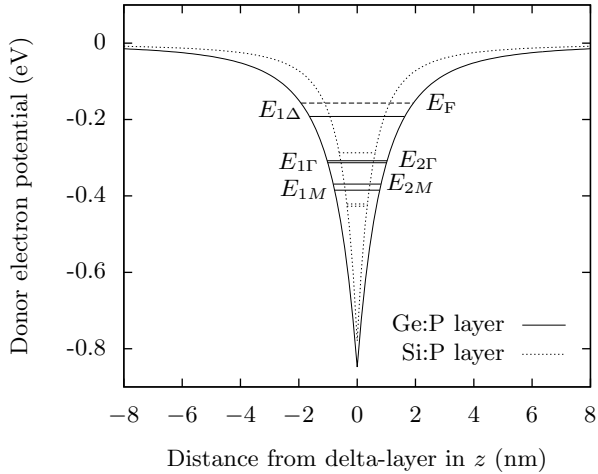}
    \caption{Donor potential for a Ge:P $\delta$-layer [solid line] and Si:P $\delta$-layer (1/4~ML) [dotted line]. The Fermi level and energies of the conduction valley minima are shown as energy levels in the potential well for the Ge:P $\delta$-layer [solids lines] and Si:P $\delta$-layer [dotted lines].}
    \label{GeP:dp}
\end{figure}

\begin{figure}[th!]
    \centering
    \includegraphics{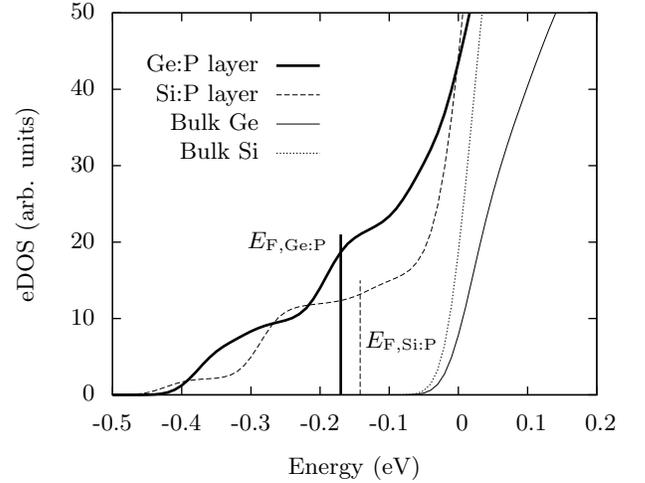}
    \caption{eDOS for the Ge:P $\delta$-layer [heavy solid line] and Si:P $\delta$-layer (1/4~ML) [dashed line], bulk Ge [light solid line] and bulk Si [dotted line]. A Gaussian smearing of 25 meV has been applied for visualisation.}
   \label{eDOS}
\end{figure}

The similar electronic properties of Ge to Si make it equally attractive for the manufacture of nanoelectronic devices. Carrying on from the success of the $\delta$-doped Si:P fabrication route, $\delta$-doped P in Ge (Ge:P) structures are now being fabricated on the nanometre scale utilising the same \textit{in situ} doping techniques.~\cite{Scappucci2011c,Scappucci2013a} Presently, experiments are focused on highly doped Ge:P $\delta$-layers~\cite{Scappucci2009a,Scappucci2011a,Scappucci2013a} which could serve as the channel material in the next generation of high-speed nano-transistors.~\cite{Kamata2008a} Ultra-scaled n-type Ge devices on the Si platform could also be realised through the patterning of Ge:P $\delta$-layers into source and drain contacts or quantum wires -- analogous to Si:P $\delta$-layers.~\cite{Weber2012a,Scappucci2012a,McKibbin2013a}

Recently fabricated Ge:P $\delta$-layers exhibit extremely high 2D carrier concentrations of 1/4~ML -- 1/2~ML,~\cite{Scappucci2011a,Scappucci2011b} which is equivalent to complete substitutional doping of the Ge(001) surface.~\cite{Scappucci2012b} However, presently there is no theoretical model for the electronic properties of these novel Ge:P nanoelectronic structures. Here, we use our TB1D model to rectify this; comparing the electronic properties of a Ge:P $\delta$-layer to that of a Si:P $\delta$-layer. There is an \textit{a priori} argument for applying the TB1D model to a Ge:P $\delta$-layer as unlike DFT,~\cite{Remediakis1999a} TB methods are able to reproduce the band structure of bulk Ge as easily as that of bulk Si (without the need for computationally expensive hybrid density functionals).

The bulk band structure of Ge is shown in Figure~\ref{ge_bulk}. Ge is an indirect band gap semiconductor and has a four-fold degenerate CB minimum at $L$ in the FCC BZ. The change from an FCC unit cell to a tP supercell projects the conduction valleys at $L$ to $M$ in the tP BZ. In Figure~\ref{ge_bands}, these bands are labelled $1M$ and $2M$. A valley splitting of 16~meV is predicted between the 1$M$ and 2$M$ band minima. The 1D confinement of the donor electrons shifts the lowest lying conduction bands into the bulk band gap region. The $1M$ and $2M$ bands are each doubly degenerate as Ge has four equivalent lowest conduction valleys. Similarly to Si, the change from a FCC unit cell to a tP supercell also folds the bands in the $\pm X$ and $\pm Y$ directions. In Figure~\ref{ge_bands}, the $\pm0.88X$ and $\pm0.88Y$ valley minima of bulk Ge are folded to $\pm0.24X$ and $\pm0.24Y$ respectively. The $\pm Z$ conduction valleys are folded to $\Gamma$. In Figure~\ref{ge_bands}, these bands are labelled $1\Gamma$ and $2\Gamma$. A valley splitting of 6~meV is predicted between the $1\Gamma$ and $2\Gamma$ band minima. Interestingly, the conduction valley at $\Gamma$ in the band structure of bulk Ge is not shifted into the bulk band gap region to the extent of the $M$ and $X$ bands. The projected $Z$ conduction valleys dominate the CB minima of the Ge:P $\delta$-layer band structure at $\Gamma$. This is explained by the confinement of the CB minima which is proportional to the curvature of the conduction valleys.~\cite{Drumm2012a} A comparison of the bands in Figure~\ref{ge_bulk} shows the conduction valley at $\Gamma$ is narrower than the bands at $M$ and $X$, and it is therefore not shifted as far into the bulk band gap region. This is shown in Figure~\ref{ge_bands} where the $\Gamma$ band of bulk Ge is above the Fermi level. Therefore, at low voltage biases we expect the transport properties of the Ge:P $\delta$-layer to be dominated by the four projected $L$ conduction valleys and two projected $Z$ conduction valleys.

\begingroup
\begin{table}[t!]
\begin{ruledtabular}
     \caption{Energies of the CB minimum, valley splitting, Fermi level and binding energy (meV) of a Ge:P $\delta$-layer for a range of ML doping densities.}
     \begin{tabular*}{\columnwidth}{@{\extracolsep{\fill} } l c c c c c c }
          ML doping density & 1/4 & 1/8 & 1/16 & 1/32 & 1/64 \\                       
          $n_{D}~(10^{14}~\rm{cm}^{-2})$ & 1.563 & 0.781 & 0.391 & 0.195 & 0.098 \\
          \hline
          $E_{1M}$ & -385 & -211 & -117 & -66 & -37 \\
          $E_{2M}-E_{1M}$ & 16 & 5 & 2 & 1 & 1 \\
          $E_{\rm{F}}$ & -170 & -61 & -36 & -26 & -15 \\
          $E_{\rm{F}}-E_{1M}$ & 215 & 150 & 81 & 40 & 22
     \end{tabular*}
     \label{GeP:dd}
\end{ruledtabular}
\end{table}
\endgroup

The donor potential for the Ge:P $\delta$-layer is shown in Figure~\ref{GeP:dp}. At the nominal doping density of 1/4 ML the larger Bohr radius of Ge results in a donor potential which is broadened in the $z$ direction but which exhibits a similar minimum value (0.85~eV) to the donor potential of the Si:P $\delta$-layer (0.78~eV). Therefore, we expect the CB minimum ($E_{1M}$) of the Ge:P $\delta$-layer to be shifted further into the bulk band gap region than the CB minimum of the Si:P $\delta$-layer. However, comparing Figures~\ref{si_bands}~and~\ref{ge_bands} the opposite is true, the Si:P $1\Gamma$ band is shifted further into the bulk band gap region than the Ge:P $1M$ band. This can be explained by the eDOS near the CB minimum. In Figure~\ref{eDOS}, there is a larger number of conduction states near the CB minimum which decreases the confinement of any one state.

Figure~\ref{eDOS} shows the eDOS of the Si:P and Ge:P $\delta$-layers.  Excluding energies in the neighbourhood of -0.25 eV, the eDOS of the Ge:P $\delta$-layer is greater than that of the Si:P $\delta$-layer on the domain (-0.4, 0.0)~eV. This is explained by the larger number of Ge:P bands or conducting modes below the Fermi level and suggests a higher conductivity for the Ge:P $\delta$-layer than the Si:P $\delta$-layer at low voltage biases. The larger eDOS for the Si:P $\delta$-layer in the neighbourhood of $E=-0.25$~eV corresponds to the filling of the Si:P $1\Delta$ band ($E_{1\Delta,\rm{Si:P}}=-287$~meV) before the Ge:P $1\Delta$ band ($E_{1\Delta,\rm{Ge:P}}=-192$~meV). The eDOS of bulk Ge and bulk Si show the extent to which the conduction valleys are confined or shifted into the bulk band gap region.

Table~\ref{GeP:dd} shows the sensitivity of CB minimum, valley splitting, Fermi level and binding energy to changes in ML doping density. Similarly to the Si:P $\delta$-layer, the CB minimum of the Ge:P $\delta$-layer ($E_{1M}$) shows a linear response to changes in areal doping density. This is not true of the Fermi level and therefore binding energy. Their non-linear behaviour can be explained by the larger number of partially filled bands below the Fermi level. Interestingly, at a doping density of 1/4 ML the value of the $2M/1M$ valley splitting is equal to more than twice the energy splitting at $\Gamma$. This suggests the $1M$ and $2M$ bands are more strongly confined than the $1\Gamma$ and $2\Gamma$ bands. This is expected as the CB minimum of bulk Ge at $L$ is 225~meV below the $X$ conduction valley minimum (see Figure~\ref{ge_bulk}). The $2M/1M$ valley splitting of the Ge:P $\delta$-layer (16~meV) is 10~meV greater than the $1\Gamma/2\Gamma$ valley splitting of the Si:P $\delta$-layer (6~meV). Therefore, we expect experimental measurements of the Ge:P $\delta$-layer to be more sensitive to the energy splitting of the lowest conduction valleys. The binding energy of the Ge:P $\delta$-state is 70~meV less than that of the Si:P $\delta$-state and we therefore expect this to be resolvable using current experimental techniques. This prediction is consistent with the larger Bohr radius of Ge.

\section{\label{sec:conclusion}Conclusions}

We find the TB1D model produces results that are comparable to more rigorous TB and DFT calculations of Si:P $\delta$-layers. A self-consistent 1D donor potential is therefore adequate to describe the physics of this highly confined and densely doped 2D system. Self-consistency is built into the semi-classical component of the donor potential. Therefore, we avoid the need to solve iteratively for the electrostatic field, increasing the computational efficiency of our calculations. The TFD formalism is known to be exact in the high density limit and here is shown to produce meaningful results at ML doping densities as low as $\sim$$2$$\times$$10^{13}~\rm{cm}^{-2}$.

The calculated values for the conduction valley minima, valley splittings and Fermi levels agree with the most recent theoretical models and experimental measurements, and are internally consistent with the calculated donor potential. We believe this validates both the TB1D model and the addition of the 1D donor potential to only the diagonal terms of the Hamiltonian matrix. XC effects are shown to be a major source of discrepancy between the models discussed herein. At the nominal 1/4~ML doping density, the XC potential for the Si:P $\delta$-layer is found to be dominated by exchange effects (in agreement with Ref.~\onlinecite{Drumm2012a}). Correlation effects are therefore unimportant in the Si:P $\delta$-layer system.

The Ge:P $\delta$-layer is shown to have conduction valley minima at $M$, $\Gamma$ and $\pm0.24X$ in the tP BZ. A Ge:P $\delta$-state, analogous to the Si:P $\delta$-state~\cite{Miwa2013a} is predicted at $M$ with a $2M/1M$ valley splitting of 16~meV. We therefore expect experimental measurements to be more sensitive to the energy splitting of the lowest conduction valleys in a Ge:P $\delta$-layer. The binding energy of this Ge:P $\delta$-state is calculated to be 70~meV smaller than the binding energy of the Si:P $\delta$-state, which suggests that this difference is experimentally resolvable using current experimental techniques. The Ge:P $\delta$-layer exhibits a larger number of partially occupied conducting modes than the Si:P $\delta$-layer. This suggests a higher conductivity for the Ge:P $\delta$-layer at low voltage biases.

\begin{acknowledgements}
The authors' thank A. Budi and D.W. Drumm for useful discussions.
\end{acknowledgements}

\appendix*
\section{\label{sec:appendix}Derivation of the electrostatic potential of the donor electrons in the Thomas-Fermi approximation}

This derivation follows in part that of March,~\cite{March1983} and Ioriatti~\cite{Ioriatti1990a} for the TF theory and TF theory of $\delta$-doped semiconductor structures. In what follows, $n$ is electron density and $z$ is a coordinate that runs perpendicular to the doping plane. SI units are used throughout.

The TF energy-density functional is written,
\begin{equation}
     E_{\textrm{TF}}[n(z)]=T[n(z)]+U_{ee}[n(z)]+U_{e\textrm{N}}[n(z)]\ , \nonumber
\end{equation}
where $T[n(z)]$ is the kinetic energy density functional and, $U_{ee}[n(z)]$ and $U_{e\textrm{N}}[n(z)]$ are the electron-electron and electron-nuclear potential energy density functionals.
\begin{equation}
     T[n(z)]=\int t\ dz\ ,
     \label{eq:big_tee}
\end{equation}
where $t$ is the kinetic energy per unit volume and can be written as
\begin{equation}
     t=\int^{p_{\rm{F}}}_{0} n(z)\ \frac{\{p(z)\}^{2}}{2\bar{m}}\ I_{z}(p)\ dp\ . \nonumber
\end{equation}
$I_{z}(p)\ dp$ is the probability of an electron having momentum between $p(z)$ and $p(z)+dp$ and can be written (to first order in $dp$) as
\begin{equation}
     I_{z}(p)\ dp=\frac{3 \{p(z)\}^2}{\{p_{\textrm{F}}(z)\}^3}\ dp\ , \nonumber
\end{equation}
where $p_{\textrm{F}}$ is the maximum momentum of an electron. Therefore,
\begin{equation}
     t=\int^{p_{\rm{F}}}_{0} n(z)\ \frac{\{p(z)\}^{2}}{2\bar{m}}\ \frac{3 \{p(z)\}^2}{\{p_{\textrm{F}}(z)\}^3}\ dp\ .
     \label{eq:little_tee}
\end{equation}
In the local density approximation,
\begin{equation}
     n(z)=n_{0}=\frac{N}{V} \nonumber
\end{equation}
where, as a consequence of the uncertainty principle,
\begin{equation}
     N=\frac{2}{h^3}\mathscr{V}\ . \nonumber
\end{equation}
$\mathscr{V}$ is the total occupied volume of phase space,
\begin{equation}
     \mathscr{V}=\frac{4\pi\nu}{3}{\{p_{\textrm{F}}(z)\}^3}V\ , \nonumber
\end{equation}
where $\nu$ is the number of equivalent conduction valleys.
Therefore,
\begin{equation}
     N=\frac{2}{h^3}\frac{4\pi\nu}{3}{\{p_{\textrm{F}}(z)\}^3}V\ , \nonumber
\end{equation}
\begin{equation}
     n(z)=\frac{8\pi\nu}{3h^3}{\{p_{\textrm{F}}(z)\}^3} \nonumber
\end{equation}
and
\begin{equation}
     \{p(z)\}^{5} = \left(\frac{3h^{3}}{8\pi\nu}n(z)\right)^{5/3}\ . \nonumber
\end{equation}
Substituting into Eq.~\ref{eq:little_tee},
\begin{align}
     \nonumber
     t&=\frac{8\pi\nu}{2\bar{m}h^{3}}\int^{p_{\rm{F}}}_{0} \{p(z)\}^{4}\ dp\\ \nonumber
     t&=\frac{8\pi\nu}{10\bar{m}h^{3}} [p_{\textrm{F}}(z)]^{5}\\ \nonumber
     t&=\frac{3}{10\bar{m}\nu^{2/3}}(3\pi^2\hbar^{3})^{2/3}\{n(z)\}^{5/3}\\ \nonumber
     t&=c_{k}\{n(z)\}^{5/3}\ ,
\end{align}
where
\begin{equation}
     c_{k} = \frac{3}{10\bar{m}\nu^{2/3}}(3\pi^2\hbar^{3})^{2/3}\ . \nonumber
\end{equation}
Finally,  the kinetic energy density functional can therefore be written as
\begin{equation}
     T[n(z)]=c_{k}\int \ n(z)^{5/3}\ dz. \nonumber
\end{equation}
The electron-electron potential energy density functional is defined as
\begin{equation}
     U_{ee}[n(z)]=\frac{e^2}{8\pi\epsilon_{r}\epsilon_{0}}\int\int\frac{n(\vec{r'})n(\vec{r})}{|\vec{r}-\vec{r'}|}\ d\vec{r}\ d\vec{r'} \nonumber
\end{equation}
and for an infinite plane of charge can be written as
\begin{equation}
     U_{ee}[n(z)]=-\frac{e^2}{4\epsilon_{r}\epsilon_{0}}\int\int n(z)\ n(z')\ |z-z'|\ dz\ dz'\ . \nonumber
\end{equation}
Similarly, the electron-nuclear potential energy density functional is defined as
\begin{equation}
     U_{eN}[n(z)]=e\int n(\vec{r})V_{\textrm{N}}(\vec{r})\ d\vec{r} \nonumber
\end{equation}
and for an infinite plane of charge can be written as
\begin{equation}
    U_{eN}[n(z)]=-\frac{e^{2}n_{D}}{2\epsilon_{r}\epsilon_{0}}\int n(z)\ |z|\ dz\ . \nonumber
\end{equation}
The TF energy-density functional is then
\begin{align}
     \nonumber
     E_{\textrm{TF}}[n(z)]=&~c_{k}\int \ n(z)^{5/3}\ dz-\frac{e^{2}n_{D}}{2\epsilon_{r}\epsilon_{0}}\int n(z)\ |z|\ dz\\ 
     &-\frac{e^2}{4\epsilon_{r}\epsilon_{0}}\int\int n(z)\ n(z')\ |z-z'|\ dz\ dz'\ .
     ~\label{eq:E_TF}
\end{align}
We seek to minimise this expression for the total energy by varying the electron density subject to a constraint, namely that the total number of electrons remains constant. This is expressed through the variational statement,
\begin{equation}
     \delta(E-\mu N)=0\ , \nonumber
\end{equation}
where $\mu$ is the Lagrangian multiplier and the chemical potential, and the total number of electrons is given by
\begin{equation}
     N=\int n(z)\ dz\ . \nonumber
\end{equation}
From this statement it follows that
\begin{equation}
     \mu=\frac{\partial{E}}{\partial{N}}\ . \nonumber
\end{equation}
Carrying out the functional derivative of Eq.~\ref{eq:E_TF} with respect to $N$,
\begin{equation}
     \mu=\frac{5}{3}c_{k}\{n(z)\}^{2/3}-\frac{e^{2}n_{D}}{2\epsilon_{r}\epsilon_{0}}|z|-\frac{e^{2}}{2\epsilon_{r}\epsilon_{0}}\int n(z)|z-z'|\ dz\ .
     \label{eq:mu}
\end{equation}
Then, defining the electrostatic part of Eq.~\ref{eq:mu} as
\begin{equation}
     V_{\rm{TF}}(z)=-\frac{e^{2}n_{D}}{2\epsilon_{r}\epsilon_{0}}|z|-\frac{e^{2}}{2\epsilon_{r}\epsilon_{0}}\int n(z)|z-z'|\ dz\ , \nonumber
\end{equation}
the self-consistency of the electrostatic field is achieved by insisting $V_{\rm{TF}}$ and $n$ are related by the Poisson equation,
\begin{equation}
     \frac{\partial^{2}}{\partial z^{2}}V_{\rm{TF}}(z)=\frac{e^{2}}{\epsilon_{r}\epsilon_{0}}n(z)-\frac{e^{2}n_{D}}{\epsilon_{r}\epsilon_{0}}\delta(z)\ .
     \label{eq:poisson}
\end{equation}
Rewriting Eq.~\ref{eq:mu},
\begin{equation}
     \mu=\frac{5}{3}c_{k}\{n(z)\}^{2/3}+V_{\rm{TF}}(z) \nonumber
\end{equation} 
and rearranging for $n(z)$,
\begin{equation}
     n(z)=\left(\frac{3}{5c_{k}}\right)^{3/2}\big(\mu-V_{\rm{TF}}(z)\big)^{3/2}\ . \nonumber
\end{equation}
We substitute this expression for $n(z)$ into Eq.~\ref{eq:poisson}, which results in the differential equation,
\begin{align}
     \nonumber
     \frac{\partial^{2}}{\partial z^{2}}V_{\rm{TF}}(z)=&~\frac{e^{2}}{\epsilon_{r}\epsilon_{0}}\left(\frac{3}{5c_{k}}\right)^{3/2}\big(\mu-V_{\rm{TF}}(z)\big)^{3/2}\\
     &-\frac{e^{2}n_{D}}{\epsilon_{r}\epsilon_{0}}\delta(z)\ .
     \label{eq:v_de}
\end{align}
The solution to equation \ref{eq:v_de} is of the form,
\begin{equation}
     V_{\rm{TF}}(z)-\mu=-\frac{\alpha^{2}}{(\alpha|z|+z_{0})^{4}} \nonumber
\end{equation}
where
\begin{equation}
     \alpha=\frac{(2\bar{m})^{3/2}e^{2}\nu}{60\pi^{2}\epsilon_{r}\epsilon_{0}\hbar^{3}} \nonumber
\end{equation}
and
\begin{equation}
     z_{0}=\left(\frac{8\alpha^{3}\epsilon_{r}\epsilon_{0}}{e^{2}n_{D}}\right)^{1/5}\ . \nonumber
\end{equation}
Finally, we set the chemical potential equal to zero, which results in the following expression for the electrostatic potential of the donor electrons,
\begin{equation}
     V_{\rm{TF}}(z)=-\frac{\alpha^{2}}{(\alpha|z|+z_{0})^{4}}\ . \nonumber
\end{equation}

\bibliography{/Users/jackson/Documents/articles/library.bib}

\end{document}